# High-Dimensional Evolutionary Algorithm Based Design of Semi-Adder

Xi Zhang, Huihui Liu , Junrui Xi , Menglu Chen and Tao Zhu

*Abstract* — Facing the physical limitations and energy consumption bottlenecks of traditional electronic devices, we propose an innovative design framework integrating evolutionary algorithms and metasurface technology, aiming to achieve intelligent inverse design of photonic devices. Based on a constructed high-dimensional evolutionary algorithm framework, a four-layer metasurface cascade regulation system was developed to realize the full optical physical expression of half-adder logic functions. This algorithm enables global optimization of 50×50×4 unit parameters and can be extended to the design of more complex functional devices,thereby promoting goal-oriented and functional customization development

*IndexTerms*—High-dimensional evolutionary algorithm; Metasurface; Cascaded regulation system; Half-adder

## I. INTRODUCTION

In recent years, the development of photonic devices has advanced significantly, and their design methodologies have reached a mature stage. However, the current mainstream forward design process remains highly reliant on the designer's experience and intuition, demanding rigorous theoretical expertise. Such approaches often yield suboptimal device structures, leading to inevitable performance compromises and technical bottlenecks. To address these limitations, inverse design has emerged as a promising paradigm for enhancing the performance and efficiency of complex photonic systems. The core principle of inverse design involves starting from predefined performance metrics or specifications and reverse-engineering the structures and parameters of systems or components required to achieve these targets. By integrating various computational intelligence algorithms—such as genetic algorithms (GA)[1], gradient descent[5]-[7], particle swarm optimization (PSO), objective-first methods, and deep learning[8]—inverse design frameworks incorporate broader parameter spaces, effectively overcoming the constraints of traditional forward design. These algorithms enable the exploration of high-dimensional solution spaces with greater precision and adaptability. Looking ahead, the integration of artificial intelligence (AI) and deep learning into the design of optical computing devices represents another transformative direction. The application of intelligent algorithms is driving the evolution of metasurface design toward goal-oriented and functionally customized architectures, which will significantly propel advancements in optical computing technologies.

Currently, researchers both domestically and internationally commonly employ heuristic intelligent algorithms to design devices, aiming to improve efficiency and performance. Examples include genetic algorithms, particle swarm optimization (PSO), and simulated annealing algorithms, which simulate natural evolutionary processes. Through iterative refinement, these methods update existing solutions and ultimately converge to an optimal result within a reasonable number of iterations. These algorithms are well-established, straightforward to implement, and possess strong global optimization capabilities, making them widely applicable in the design of micro- and nano-photonic devices. For instance, in 2021, Hongyi Yuan developed a hybrid optimization algorithm combining simulated annealing and genetic algorithms to design cascaded bandpass filters and wavelength routers[2]. The resulting bandpass filter exhibited a 408 nm operational bandwidth with a transmission rate exceeding 80%. Similarly, in 2023, Miao et al. integrated deep neural networks with genetic algorithms to create a deep learning model[3] for the inverse design of 2D phononic crystal dispersion relations.

However, despite the significantly higher training efficiency of artificial neural networks (ANNs)[4] compared to evolutionary algorithms in inverse design when sufficient datasets are available, their practical application still faces substantial challenges. Firstly, ANNs encounter the non-unique mapping problem in photonics inverse design tasks. Specifically, during the training process for inverse design, a single electromagnetic response may correspond to multiple structural configurations. When such conflicting data exist in the training dataset, it becomes extremely difficult to train a convergent neural network. Secondly, during real-world simulations, datasets with adequate sample sizes for training are often unavailable.Although ANNs demonstrate superior inverse design efficiency over evolutionary algorithms when sufficient data is provided, their practical application in photonics optimization remains constrained by two critical limitations. First, the inherent one-to-many correspondence between structural configurations and electromagnetic responses introduces contradictory patterns into the training dataset, fundamentally hindering the convergence of neural networks. Second, real-world scenarios are plagued by severe data scarcity, as generating comprehensive electromagnetic response libraries through full-wave simulations imposes insurmountable computational costs for large-scale tasks. These dual challenges collectively undermine the reliability and generalization capability of ANN-based inverse design frameworks in nanophotonics.

In this letter, Building upon the concept of inverse design, we address the limitations of evolutionary algorithms in solving high-dimensional problems by developing a high-dimensional evolutionary optimization algorithm capable of achieving global optimization for complex systems. This algorithm is specifically tailored for ultra-large-scale parameter spaces and incorporates a meticulously designed objective function to ensure high-quality outputs across all input combinations.

## II. RESULTS AND DISCUSSION

A. Construction of the Algorithm

To address the inefficiency of traditional evolutionary algorithms in handling ultra-high-dimensional parameter spaces, we propose an evolutionary algorithm framework tailored for ultra-high-dimensional electromagnetic inverse design. This algorithm is capable of effectively managing ultra-large-scale parameter spaces while maintaining global ptimization capabilities and significantly enhancing computational efficiency. The proposed algorithm adopts a hybrid strategy combining hierarchical encoding and adaptive operations. Its core architecture includes the following key components: hierarchical chromosome encoding, adaptive gene recombination strategy, and distributed population management.

First, the traditional single chromosome structure is divided into a multi-layer hierarchical encoding, where each layer represents the phase distribution of a metasurface. This hierarchical encoding scheme enables the algorithm to simultaneously optimize the parameters of multiple metasurface layers, significantly enhancing the scalability of the parameter space. To address the curse of dimensionality in ultra-high-dimensional spaces, the algorithm incorporates an adaptive gene recombination mechanism based on information entropy. This mechanism dynamically adjusts the probabilities and ranges of crossover and mutation operations, ensuring a balance between exploration and exploitation. Specifically, when the algorithm becomes trapped in a local optimum, the mutation rate is automatically increased to escape suboptimal regions. Conversely, when promising areas are identified, the local search intensity is enhanced to refine the solution further. Finally, a distributed population management strategy based on the "island model" is employed. The overall population is divided into multiple subpopulations, each evolving independently within different parameter subspaces. Periodic migration mechanisms allow the exchange of elite individuals between subpopulations, maintaining population diversity while preventing premature convergence.

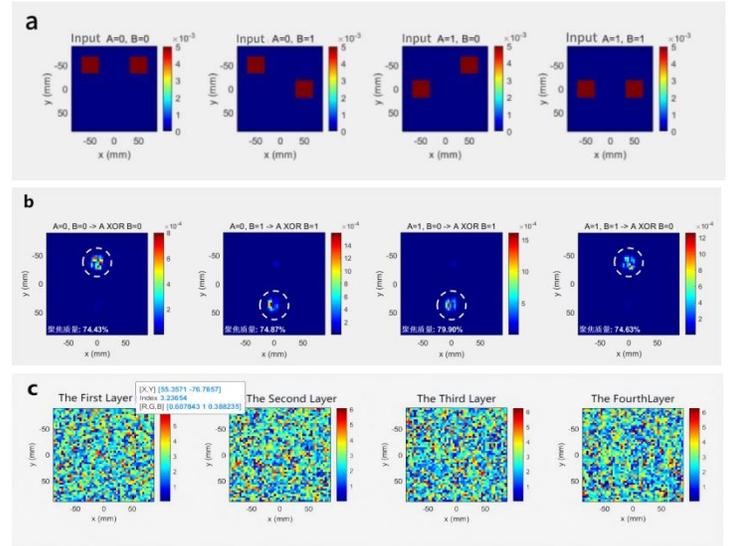

Fig.2. (a) Schematic diagram of the input port. (b) Schematic diagram of the output port. (c) Schematic diagram of the output port.

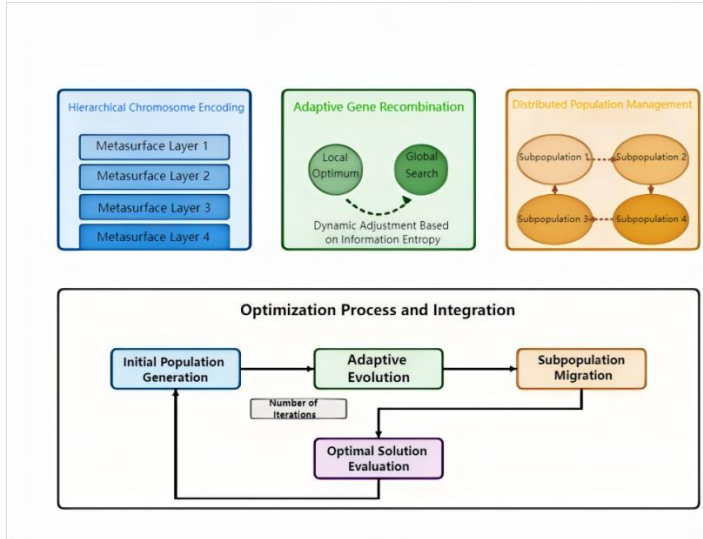

Fig.1. Algorithm Architecture Diagram

B. Design of the Objective Function

A multi-objective weighted fitness function was designed based on the logical requirements of the half-adder. This function simultaneously considers the output accuracy, contrast, and system energy efficiency under four input combinations ([0,0], [0,1], [1,0], [1,1]). The fitness function is defined as follows:

$$F = -\sum_{n=1}^{N} \left( \frac{\int_{\Omega_T} |E_n(x,y)|^2 dxdy}{\int_{all} |E_n(x,y)|^2 dxdy} - \alpha \frac{\int_{\Omega_{NT}} |E_n(x,y)|^2 dxdy}{\int_{all} |E_n(x,y)|^2 dxdy} \right)$$

Here, nT represents the field in the target region. The following are the simulation results based on this algorithm implemented in MATLAB.

This algorithm demonstrates significant advantages over traditional evolutionary algorithms: Through hierarchical encoding and parallel computing strategies, the convergence speed of this algorithm is two orders of magnitude faster than that of a standard genetic algorithm, demonstrating exceptional performance in handling parameter spaces on the order of tens of thousands of dimensions. The adaptive operation mechanism enables the algorithm to dynamically adjust its search strategy, effectively preventing premature convergence and enhancing its robustness in solving complex problems. The hierarchical encoding and distributed population architecture provide excellent scalability, allowing the algorithm to easily accommodate the growth of parameter scales. This lays a solid algorithmic foundation for the design of more complex functional devices, such as full adders. Through carefully designed population migration strategies and mutation mechanisms, the algorithm maintains strong global optimization capabilities, effectively avoiding entrapment in local optima. In practical implementation, the algorithm employs a population size of 100 individuals and runs for 5000 iterations. Through fine-tuned parameter adjustments, high-quality design solutions are achieved within reasonable computational resources. Using this algorithm, a four-layer metasurface system (with 50×50 units per layer) successfully realizes the precise phase distributions required for half-adder functionality, providing a theoretical foundation for subsequent unit cell structure design and system implementation. Preliminary tests indicate that the algorithm achieves a success rate of over 95% in 50 independent runs, with an average convergence generation of approximately 3000. Compared to traditional algorithms solving problems of similar scale, the efficiency is improved by approximately 200 times, fully demonstrating its outstanding performance in optimizing ultra-high-dimensional parameter spaces.

C. Array Construction and Full-Wave Simulation

According to the generalized Snell's law[9], the most critical aspect of achieving wavefront modulation with metasurfaces is introducing abrupt phase changes at the interface to generate a phase gradient. Previous studies have proposed various micro- and nanostructure units capable of providing a complete 0–2π phase range. Currently, the mechanisms for phase modulation in micro- and nanostructures are mainly categorized into three types: resonant phase[10], propagation phase[11], and geometric phase[12]. The modulation mechanism employed in this project is geometric phase. Below, we introduce the generation mechanism of geometric phase and briefly outline the design process for wavefront-modulating metasurfaces. Geometric phase, also known as Pancharatnam-Berry (P-B) phase, is a method of introducing abrupt phase changes that is fundamentally different from resonant phase and propagation phase. Geometric phase arises from an adiabatic cyclic process and can be achieved by rotating structural units. Based on Huygens-Fresnel principle, when electromagnetic waves propagate in free space, the electromagnetic field distribution of any wavefront can be considered as the superposition of radiation waves from secondary point sources on the previous wavefront. Leveraging the Huygens-Fresnel principle, metasurfaces can be designed to achieve various wavefront modulation functions.

After the target phase is computed using the algorithm, this study tests the design based on a classic resonator unit. Below, we introduce its geometric structure and electromagnetic characteristics. The unit consists of a rectangular substrate and a split-ring structure. By adjusting the opening angle of the split ring, the phase of the electromagnetic wave can be modulated.

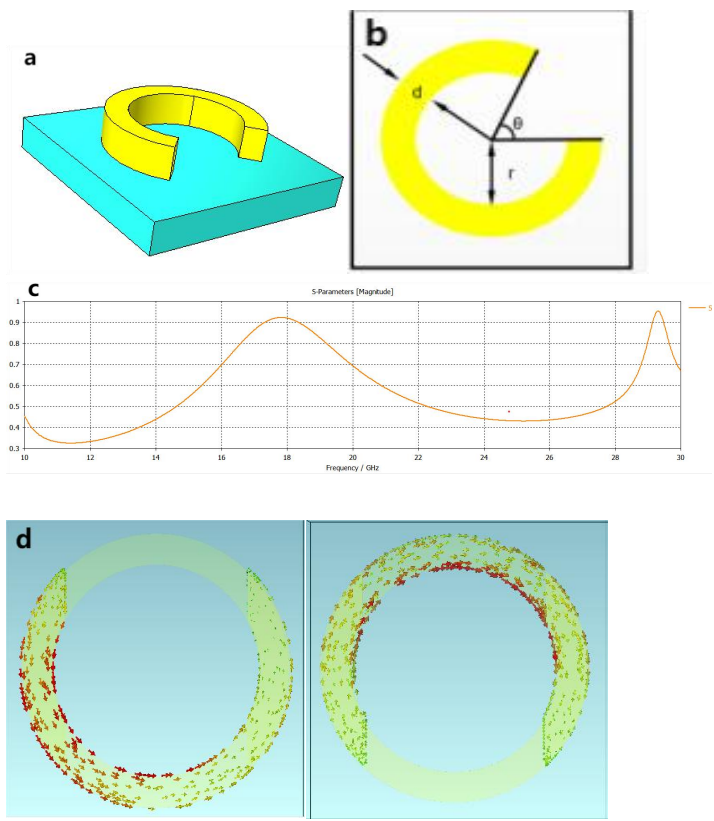

Fig.3. (a)Resonator ring unit model and parameters.(b)S-parameter characteristics.(c)Surface current distribution

We employ a combined simulation approach using Matlab and CST (CST Studio Suite), which facilitates rapid modeling and enables accurate evaluation of system performance. Based on the existing unit cell structure, we proceed to construct a complete model of the metasurface.

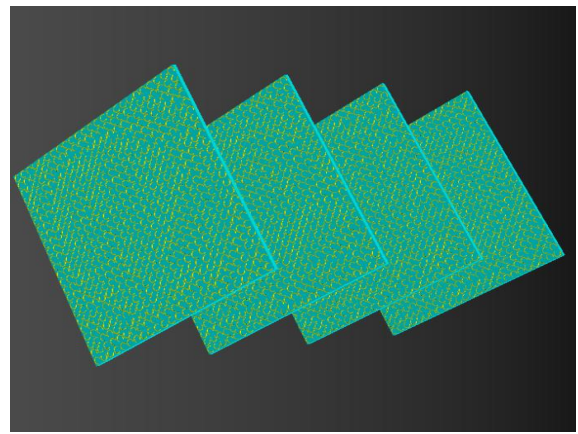

Figure.3. Four-layer cascaded metasurface

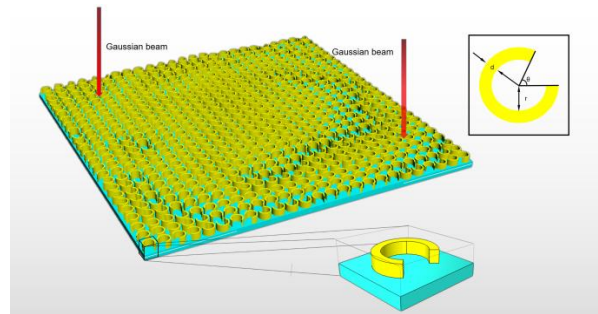

Figure.4. Single-layer surface structure distribution

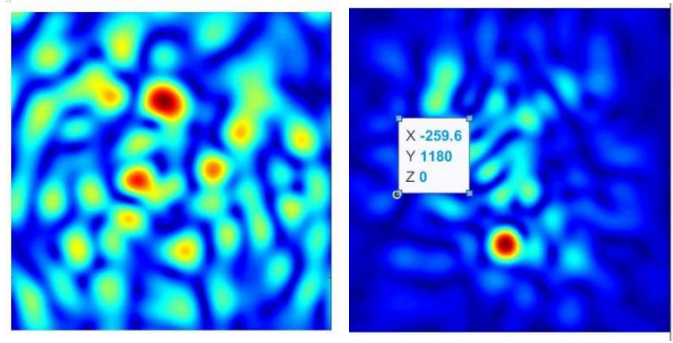

Figure.5. Focusing effect

The left and right images clearly demonstrate an optimized focusing effect, where most regions exhibit low field intensity (blue), while a distinct high-intensity hotspot (red-yellow region) is observed at specific locations. This precise field intensity focusing represents one of the key objectives of the algorithm optimization, corresponding to the output state of the half-adder under a specific input combination . These simulation results validate that the ultra-high-dimensional evolutionary algorithm designed in this project can effectively optimize the structure of multi-layer metasurfaces, successfully achieving the desired electromagnetic field modulation. The algorithm enables precise energy focusing at specific locations, thereby realizing the logical functionality of the half-adder.

III. CONCLUSION

This study proposes an intelligent inverse design framework that integrates evolutionary algorithms with metasurface technology for the innovative design of photonic devices. To address the limitations of traditional forward design methods, such as restricted geometric complexity, cumbersome parameter scanning, and suboptimal solutions, we have developed a high-dimensional evolutionary optimization algorithm tailored for large-scale electromagnetic inverse design. This work establishes a cross-paradigm between evolutionary computation and nanophotonics, offering a novel approach for goal-oriented optical computing device design. It lays the foundation for constructing complex arithmetic logic units in all-optical processors and holds significant potential for advancing high-energy-efficiency computational architectures.

REFERENCES

[1] 颜树华. 衍射微光学设计.(第 1 版).北京: 国防工业出版社,2011. 66~136.
[2] Yuan H,Ma L,Yuan Z,et al.On-Chip Cascaded Bandpass Filter and Wavelength Router Using an Intelligent Algorithm[J].IEEE Photonics Journal,2021,13(4): 6600408.
[3] MIAO X B，DONG H W，WANG Y S．Deep learning of dispersion engineering in two-dimensional phononic crystals［J］Engineering Optimization，2023，55( 1) : 125-139
[4] JAIN A, MAO J, MOHIUDDIN K. Artificial neural networks: a tutorial[J/OL]. Computer, 1996, 29(3):3144. DOI: 10.1109/2.485891.
[5] Bendsøe M P, Sigmund O. Topology Optimization: Theory, Methods and Applications. Springer, 2003.
[6] Sethian J A. Level Set Methods and Fast Marching Methods. Cambridge, 1999.
[7] Jensen J S, Sigmund O. Topology optimization for nano-photonics. Laser Photon. Rev., 2011, 5: 308-321.
[8] LeCun Y, Bengio Y, Hinton G. Deep learning. Nature, 2015, 521: 436-444
[9] N. Yu, P. Genevet, M. A. Kats, et al. Light propagation with phase discontinuities:Generalized laws of reflection and refraction. Science, 2011, 334: 333~337.
[10] E. Nazemosadat, M. Mazur, S. Kruk, et al. Dielectric broadband metasurfaces for fiber mode-multiplexed communications. Advanced Optical Materials, 2019, 7 (14): 1801679.
[11] M. Khorasaninejad, A. Y. Zhu, C. Roques-Carmes, et al. Polarization-insensitive metalenses at visible wavelengths. Nano Letters, 2016, 16(11): 7229~7234.
[12] J. P. B. Mueller, N. A. Rubin, R. C. Devlin, et al. Metasurface polarization optics: independent phase control of arbitrary orthogonal states of polarization. Physical Review Letters, 2017, 118 (11): 113901.